\def\prg#1{\medskip{\bf #1}}
\def\lra{\leftrightarrow}        \def\Ra{\Rightarrow}
               \def\pd{\partial}

\def\ess{{\mbox{\scriptsize ess}}}

\def\bR{{\bar R}}     
     
\def\bA{{\bar A}}     \def\bt{{\bar\tau}}   \def\bl{{\bar\lambda}}
\def\bth{{\bar\theta}}     
   
          \def\bb{{\bar b}}
     
\def\cL{{\cal L}}     \def\cM{{\cal M }}    \def\cK{{\cal K}}
\def\cO{{\cal O}}          \def\cB{{\cal B}}
\def\tG{{\cal G}}     \def\tK{{\cal K}}     \def\cA{{\cal A}}

\def\m{\mu}           \def\n{\nu}           \def\k{\kappa}
        \def\g{\gamma}        \def\d{\delta}
\def\S{\Sigma}                \def\t{\tau}
\def\a{\alpha}        \def\b{\beta}         \def\th{\theta}
\def\vphi{\varphi}    \def\ve{\varepsilon}  
\def\p{\pi}           \def\r{\rho}          
\def\L{\Lambda}       \def\l{\lambda}       
\def\Om{\Omega}         
         \def\bphi{{\bar\phi}}

\def\bull{\raise.15ex\hbox{\vrule height.8ex width.8ex}}
\def\nn{\nonumber}
\def\be{\begin{equation}}             \def\ee{\end{equation}}
\def\ba#1{\begin{array}{#1}}          \def\ea{\end{array}}
\def\bea{\begin{eqnarray} }           \def\eea{\end{eqnarray} }
\def\beann{\begin{eqnarray*} }        \def\eeann{\end{eqnarray*} }
\def\beal{\begin{eqalign}}            \def\eeal{\end{eqalign}}
\def\lab#1{\label{eq:#1}}             \def\eq#1{(\ref{eq:#1})}
\def\bsubeq{\begin{mathletters}}      \def\esubeq{\end{mathletters}}
\def\bitem{\begin{itemize}}           \def\eitem{\end{itemize}}

\documentstyle[aps,preprint,eqsecnum]{revtex}      

\begin{document}
\tighten


\title{On the classical central charge}

\author{M.\ Blagojevi\'c and M. Vasili\'c\thanks{Email
        addresses:  mb@phy.bg.ac.yu, mvasilic@phy.bg.ac.yu}}
\address{Institute of Physics, P.O.Box 57, 11001 Belgrade, Serbia}
\maketitle

\begin{abstract}
In the canonical formulation of a classical field theory, symmetry
properties are encoded in the Poisson bracket algebra, which may have a
central term. Starting from this well understood canonical structure,
we derive the related Lagrangian form of the central term.
\end{abstract}


\section{Introduction}

In the canonical description of a classical field theory, the
symmetry generators are represented by the phase-space functionals
which act on the basic dynamical variables through Poisson brackets
(PBs). As a consequence, the commutator algebra of the symmetry
transformations is represented by the PB algebra, which may have
central extension. If the symmetry group is parametrized by $\ve^a$,
and $G=G[\ve]$ is the canonical generator, the PB algebra has the
general form \cite{1}
\be
\{G[\ve],G[\eta]\}=G[\th]+K[\ve,\eta] \, ,                 \lab{1.1}
\ee
where $\th=\th(\ve,\eta)$ is defined by the group composition law,
and $K[\ve,\eta]$ is a constant phase-space functional---the
classical {\it central term}.

The construction of well defined (finite and differentiable)
canonical generators $G[\ve]$ has been thoroughly treated in the
Hamiltonian formalism \cite{2,3,4,5,6}. In the first step, one adopts
specific asymptotic conditions on the canonical variables, so as to
restrict the phase space to the domain on which the canonical
generators are well defined. The symmetry group itself is thereby
restricted to a subgroup which preserves the asymptotics---the {\it
asymptotic symmetry group\/}. In the second step, the asymptotic
symmetry generators are improved by adding suitable surface terms,
whereupon the PB algebra \eq{1.1} naturally emerges.

The adopted asymptotics, defined by the sole requirement of {\it
finiteness\/} and {\it differentiability\/}, is necessary for the
algebra \eq{1.1} to make sense, but it does not guarantee the
conservation of the related charges---the on-shell values of the
asymptotic symmetry generators. In what follows, we shall restrict our
interest to finite energy field configurations of isolated physical
systems, as their asymptotic behavior ensures the {\it conservation of
the canonical charges\/}.

The nature and origin of central terms in algebras associated with
physically relevant symmetries is of particular importance for a
proper understanding of the quantum features of gravity
\cite{7,8,9,10,11,12}. Central terms in quantum physics do not
emerge exclusively as a consequence of the quantum ordering
procedure, as in string theory with its Virasoro algebra, but also
inherit the underlying classical structure. In the conformal field
theory, for instance, in Cardy's formula for the asymptotic density
of states, the dominant contribution involves the classical central
term. The same property holds also for the Bekenstein-Hawking entropy
of the BTZ black hole. Thus, classical central terms are as
physically relevant as their quantum counterparts. The mechanism by
which a classical feature (such as classical central term), defined
entirely in the realm of classical physics, influences the nature of
the quantum theory, is not yet sufficiently clear and deserves
further investigation. In addition to this, classical central terms
are also related to certain geometric properties of the classical
solutions.

Although the structure of central term is well understood at the
canonical level, the whole formalism is somewhat indirect: in order
to find out the PB algebra with central term, one needs a complete
information regarding its Hamiltonian structure (Hamiltonian,
canonical generator, etc.). To circumvent this unpractical side of
the approach, one would like to find a counterpart of these results
within the Lagrangian formalism, which will eventually lead us to the
same information in a more direct way.

In the Lagrangian approach to the classical central term, one is
effectively faced with the problem of finding the counterpart of the
PB algebra \eq{1.1}. To this end, we note that this relation can be
rewritten as
\be
\d_{\eta}G[\ve]=G[\th]+K[\ve,\eta] \, ,                    \lab{1.2}
\ee
where $\d_{\eta}G[\ve]$ is the infinitesimal symmetry transformation of
$G[\ve]$ with parameter $\eta$. As the on-shell values of $G[\ve]$
represent the canonical conserved charges, it is natural to expect that
the transformation properties of the Lagrangian conserved charges may lead
us to the Lagrangian counterpart of the canonical central term
$K[\ve,\eta]$.

A number of authors in the literature have taken this way to the
problem \cite{13,14,15,16,17}. They make use of the fact that continuous
symmetries of the classical action lead, via Noether's theorem
\cite{18}, to differentially conserved Noether currents. However,
Noether current is not defined uniquely, but only up to an exact
term, the exterior derivative of the so-called superpotential. In
gauge theories the complete conserved current reduces on-shell merely
to the superpotential term. Since the superpotential is completely
arbitrary, the related Noether charge is essentially undefined. In
order to define the Noether charge, one has to find an appropriate
criterion which will fix the superpotential. As it turns out, this
criterion essentially boils down to the very existence and
conservation of Noether charges \cite{15}.

Although many important features of the Lagrangian approach have been
successfully developed, there are still some open questions; in
particular, its equivalence with the Hamiltonian formalism has not
been fully understood. The purpose of the present paper is to define
central term of the asymptotic symmetry in the Lagrangian context,
starting from the related canonical structure \eq{1.1}. The approach
leads to a very simple but quite general Lagrangian expression for
the central term, valid in an arbitrary field theory. In the case of
gauge theories, the central term takes the form of a surface
integral, as expected. The above procedure represents the first step
towards a complete Lagrangian treatment of the asymptotic symmetries,
fully equivalent with the the canonical structure expressed in
\eq{1.1}.

The layout of the paper is as follows. In Sect. II, we review some
basic aspects of the central term in the canonical formalism, including
the canonical origin of this term, and some of its basic properties. In
Sect. III, we develop an equivalent Lagrangian description of the
central term, leading to an explicit Lagrangian expression for
$K[\ve,\eta]$. In Sect. IV we apply these results to gauge theories; in
particular, we represent central term in the form of a surface
integral. In Sect. V we study several interesting examples:
Chern-Simons theory, Liouville theory and general relativity. Section
VI is devoted to concluding remarks, while some technical details are
relegated to the appendix.

\prg{Basic concepts and conventions.} Before continuing with the main
exposition, we present here a brief account of the basic concepts and
conventions used in our paper.

(1) We consider a generic classical field theory defined on an
$n$-dimensional manifold $\cM$, conventionally called spacetime. Basic
dynamical variables are the fields $\phi=\{\phi^i,\,i=1,...,m\}$
defined on $\cM$; they can be scalars, gauge fields, metric tensor etc.
The topology of spacetime $\cal M$ is assumed to have the form
$R^1\times \S$, where $\S$ is referred to as the spatial section of
$\cal M$. When the gravitational interaction is neglected, the
$n$-dimensional spacetime has the structure of Minkowski space, but in
general, the structure of $\cM$ may be quite different. The fields are
subject to the field equations derived from an action functional---the
integral of a Lagrangian $n$-form $\cL(\phi,\pd\phi)d^nx$, where the
Lagrangian function $\cL(\phi,\pd\phi)$ is a scalar density.

(2) In addition to the field equations, we impose a set of asymptotic
(or boundary) conditions, which restrict the allowed configurations of
fields in the asymptotic region. Typically, the asymptotic region is
defined as the spatial infinity. In more general situations, the
asymptotic region may be different, or even missing (see our second
example). Typical asymptotic conditions in gravity correspond to the
assumption of flat spacetime structure at spatial infinity, but our
approach allows for more general situations, such as anti-de Sitter
asymptotics, for instance. In fact, the only limitations on the type of
asymptotic behaviour that we need are some quite general requirements
on the structure of symmetry generators, as we shall see.

(3) The field equations of our concern are invariant under the action
of a continuous group of symmetries $U$ with parameters
$\ve=(\ve^a)$. The group $U$ may be a Lie group $(\ve^a$ = const.),
but it may also have a more general structure, as in gauge field
theories or gravity. Once we adopt the set of asymptotic conditions as
an additional element of the dynamical structure of the theory, the
original symmetry group $U$ is accordingly modified: it is reduced to
the subgroup $U_0\subseteq U$, consisting of all those
transformations that do not change the asymptotic conditions. In this
way, the asymptotic conditions on dynamical variables induce the
related conditions on the group parameters $\ve^a$.

(4) In the canonical formalism, basic dynamical variables are the
functions $\phi(x),\pi(x)$ on $\S$, which are the elements of the phase
space. By construction, the canonical generator of the symmetry
transformations belonging to $U_0$ is a local functional,
$G[\ve,\phi,\pi]=\int_\S d^{n-1}x\, {\cal
G}(\ve,\pd\ve,\phi,\pd\phi,\pi,\pd\pi)$. When $U_0$ is a Lie group, we
have $G[\ve,\phi,\pi]=\ve^aG_a[\phi,\pi]$. The canonical generator acts
on basic dynamical variables via the Poisson bracket operation, which
is defined in terms of functional derivatives.

Consider a local functional $F[\phi,\pi]=\int_\S
d^{n-1}xf(\phi,\pi;\pd\phi,\pd\pi)$, where the functions
$\phi(x),\pi(x)$ are defined by certain smoothness properties,
asymptotic (or boundary) conditions, and possible additional
restrictions \cite{17}. The specific form of these characteristics
depends on the theory we are interested in, and the specific problem we
wish to study. The phase space is the space of functions $\Phi: \S\to
R^{2m}$ subject to these additional conditions. A well founded
mathematical analysis of the functionals $F[\phi,\pi]$ demands a
precise definition of the space $\Phi$, which turns out, in general, to
be a rather complicated task (the interested reader may consult
\cite{19}, for instance).

The functional $F[\phi,\pi]$ is said to have well defined functional
derivatives if its variation $\d F\equiv
F[\phi+\d\phi,\pi+\d\pi]-F[\phi,\pi]$ can be written in the form
$$
\d F[\phi,\pi]=\int_\S d^{n-1}x\left[A\d\phi +B\d\pi \right]\, ,
$$
where derivatives of $\d\phi$ and $\d\pi$ do not appear on the right-hand
side \cite{2,4}.

Although the canonical generator $G[\ve,\phi,\pi]$ does not satisfy
these requirements in general, its form can be improved by adding a
suitable surface term, whereupon it becomes differentiable. The
procedure depends essentially on the form of asymptotic conditions. In
the present discussion, explicit form of the asymptotic conditions is
not needed. Instead, we assume that the asymptotic conditions are
chosen so as to guarantee differentiability, finiteness and
conservation of the improved canonical generators (see section II).

In what follows, we shall use the notation $G[\ve,\phi,\pi]$ for the
improved generator, and moreover, the functional dependence on $\phi$
and $\pi$ will be often omitted for simplicity, reducing the complete
notation $G[\ve,\phi,\pi]$ just to $G[\ve]$.

(5) Asymptotic conditions are an intrinsic part of the canonical
formalism, as they define the phase space in which the canonical
dynamics takes place. Since these conditions are invariant under the
action of the subgroup $U_0$ of $U$, it seems natural to take $U_0$ for
the asymptotic symmetry group of the theory. However, in gauge theories
and gravity, $U_0$ contains the set of residual (or pure gauge)
transformations, which are defined by the property that their canonical
generators weakly vanish. These transformations are nontrivial
``inside" ${\cal M}$, but act trivially in the asymptotic region.
Consequently, they are irrelevant for our discussion of the conserved
charges and central terms, both of which are given as some surface
integrals. Thus, we are naturally led to introduce the improved
definition of the asymptotic symmetry: it is the symmetry defined as
the factor group of $U_0$ with respect to the group of residual
symmetries \cite{1}.

(6) Our conventions are as follows. In the general analysis, the Latin
indices $(i,j,k,...)$ are used to count field components, and
$(a,b,c,...)$ are group indices. The Latin indices $(i,j,k,...)$, when
used in examples, refer to the local Lorentz frame, the Greek indices
$(\m,\n,\r,...)$ refer to the coordinate frame, and both run over
$0,1,...,n-1$; the Greek indices $(\a,\b,\g,...)$ are the space
indices, and run over $1,2,...,n$;  spacetime derivatives are often
denoted by comma: $\pd_{\m}\phi\equiv\phi_{,\,\m}$; totally
antisymmetric tensor $\ve^{ij...k}$ and the related tensor density
$\ve^{\m\n...\r}$ are both normalized by $\ve^{01...n-1}=+1$;
$\eta_{\m\n}=(+,-,...,-)$ are components of the Minkowski metric, and
$\Box=\eta^{\m\n}\pd_{\m}\pd_{\n}$.

\section{Canonical analysis}

In the canonical description of a classical field theory, the basic
dynamical variables are the phase-space coordinates $(\phi^i,\pi_i)$.
If the theory possesses a symmetry, the equations of motion are
invariant under the transformations
\be
\d_\ve\phi^i\equiv \{\phi^i,G[\ve]\}\, ,\qquad
\d_\ve\pi_i\equiv \{\pi_i,G[\ve]\}\, ,                     \lab{2.1}
\ee
where the canonical generators are phase-space functionals,
$G[\ve]\equiv G[\ve,\phi,\pi]$, which act on the phase-space
variables through the PB operation. They are finite and
differentiable functionals under appropriate asymptotic conditions,
which restrict the original symmetry to the so-called asymptotic
symmetry \cite{2,4}. The group structure of the symmetry
transformations is expressed by the closure of their commutator
algebra:
\be
[\d_\eta,\d_\ve]X=\d_\th X\, ,                             \lab{2.2}
\ee
where  $X=X(\phi,\pi)$, and $(\ve,\eta)\to\th(\ve,\eta)$ is the
composition law of the group.

\prg{The PB algebra.}
It is clear that adding a constant phase-space functional $c[\ve]$ to
the generator $G[\ve]$ will not change its action on the phase space.
Therefore, it is not really the phase-space functionals $G[\ve]$ which
are in one-to-one correspondence with the symmetry transformations, but
their equivalence classes, defined by
$$
G[\ve]\sim G[\ve])+c[\ve]\, .
$$
This fact has serious consequences on the structure of the underlying
canonical symmetry algebra. Indeed, the group structure \eq{2.2} of the
symmetry transformations implies
$$
\{X,\{G[\ve],G[\eta]\}\}=\{X,G[\th]\} \, ,
$$
which guarantees the closure of the PB algebra of the corresponding
equivalence classes of generators, while the canonical generators
themselves, in general, satisfy the PB algebra with central term,
\be
\{G[\ve],G[\eta]\}=G[\th] + K[\ve,\eta]\, ,                \lab{2.3}
\ee
where $K[\ve,\eta]$ is constant over the whole phase space. In
particular, if we restrict ourselves to a rigid internal symmetry
group, we have $\th^c=f_{ab}{^c}\ve^a\eta^b$, and the above algebra
takes the simpler form:
$$
\{G_a,G_b\}=f_{ab}{^c}G_c + K_{ab}\, ,
$$
where we used $G[\ve]=\ve^aG_a$, $K[\ve,\eta]=\ve^a\eta^b K_{ab}$. In
what follows, we shall address the important question of how and to
what extent the classical central term is related to the structure of
a given classical theory.

\prg{Asymptotic conditions.} Since the canonical generators act via
the PB operation, they must be both finite and differentiable
functionals \cite{2,4,5}. This property is ensured by a proper
construction of $G[\ve]$ and appropriate boundary conditions at spatial
infinity, which restrict the allowed values of both the phase-space
variables and the symmetry parameters $\ve$. For any two generators
$G[\ve]$ and $G[\eta]$, their PB algebra is a realization of the
structure of the asymptotic symmetry group, and has the general form
\eq{2.3}, where $K[\ve,\eta]$ is the canonical central term \cite{1}.
Thus, the necessary asymptotic conditions for the algebra \eq{2.3} to
make sense are defined by the requirements of
\bitem
\item[$(i)$] finiteness and differentiability of the generators.
\eitem
However, these conditions are not sufficient to ensure the
conservation of the related charges. Indeed, every symmetry generator
must obey the relation \cite{3,6}
\be
\frac{\pd G}{\pd t} + \{ G,H_T \} = C_{PFC} \, ,           \lab{2.4}
\ee
where $H_T$ is the total Hamiltonian, and $C_{PFC}$ stands for
primary first-class constraints. This relation holds only up to trivial
generators, such as constants or surface integrals. Since the temporal
change of $G$ is given by the formula
$$
\frac{dG}{dt}=\frac{\pd G}{\pd t}+ \{G,H_T\} \, ,
$$
the weak equality $dG/dt \approx 0$ will hold only if the generators
are redefined to absorb these constants, and the problematic surface
terms are eliminated by the additional asymptotic conditions. In what
follows, we shall further restrict the asymptotics by the requirement
of
\bitem
\item[$(ii)$] the conservation of energy.
\eitem
This will, in addition, imply the conservation of all the canonical
charges---the property we need for the consistent Lagrangian
treatment of central charges.

\prg{Trivial central terms.} We have already seen that additive
constants in the canonical generators do not influence the symmetry
transformation law \eq{2.1}. However, as seen from \eq{2.3}, such a
change of $G[\ve]$ induces a change of the central term:
\be
G[\ve]\, \rightarrow\, G[\ve] + c[\ve] \quad\Rightarrow\quad
K[\ve,\eta]\, \rightarrow\, K[\ve,\eta] - c[\th] \, .      \lab{2.5}
\ee
Conversly, if a central term has the form $K[\th(\ve,\eta)]$, it is
always possible to redefine the generators according to
$G[\ve]\rightarrow G[\ve] + K[\ve]$ so that the new central term
vanishes.
\bitem
\item[(1)]{Any central term which can be eliminated by the simple
redefinition of symmetry generators $\,G[\ve]\rightarrow G[\ve]
+c[\ve]$, is considered trivial}.
\eitem
We shall often ignore trivial central terms, as they can always be
transformed to zero.

There are two particular cases in which one can immediately recognize
trivial central terms \cite{1,15}. The first is defined by the
existence of at least one phase-space point invariant under the
action of the symmetry group. Indeed, if $\phi^i$ and $\pi_i$ have
vanishing PBs with all the generators $G[\ve]$ at some point
$(\bar\phi,\bar\pi)$, then $\pd G[\ve]/\pd\phi^i$ and $\pd
G[\ve]/\pd\pi_i$ vanish at $(\bar\phi, \bar\pi)$. As a consequence,
the PB $\{G[\ve],G[\eta]\}$ also vanishes, and the central term takes
the form $K[\ve,\eta]=-G[\th,\bar\phi,\bar\pi]$. Therefore,
\bitem
\item[(2a)]{if there exists a phase-space point invariant under the
action of the symmetry group, the corresponding central term is
necessarily trivial}.
\eitem
The second case refers to symmetry groups whose group structure does
not allow for nontrivial central extensions. Indeed, applying the
Jacoby identity to \eq{2.3}, we obtain relations which restrict the
values of $K[\ve,\eta]$:
\[
K[\,\th(\ve,\eta),\t\,] + \mbox{cyclic}\,(\ve,\eta,\t)=0\,.
\]
As a consequence, it may happen that this constraint has no
non-trivial solutions for $K[\ve,\eta]$. Such an
example is described by the following theorem:
\bitem
\item[(2b)]{If the algebra of symmetry generators is a semi-simple
Lie algebra, the central term is necessarily trivial}.
\eitem
The proof is based on the observation that the constraint
$f_{ab}{^d}K_{dc} +\mbox{cyclic}\,(a,b,c)=0$ has no nontrivial
solutions for $K_{ab}$ if the structure constants $f_{ab}{^c}$
define a semi-simple Lie algebra.

\prg{Canonical equivalence.} Given a Hamiltonian classical theory,
one can always redefine phase-space variables to obtain canonically
equivalent theory. This is done by means of canonical
transformations,
\[
\phi \rightarrow \phi' = \phi'(\phi,\pi)\,,\qquad
\pi \rightarrow \pi' = \pi'(\phi,\pi)\,,
\]
defined by demanding the invariance of the PB structure:
$$
\{A',B'\}' = \{A,B\}
$$
for every $A'\equiv A'(\phi',\pi')=A(\phi,\pi)$, and similarly for
$B'$. The symmetry generators of the canonically equivalent theory
are simply $G'[\ve,\phi',\pi']= G[\ve,\phi,\pi]$, and therefore,
\be
\{G'[\ve],G'[\eta]\}'=\{G[\ve],G[\eta]\}=G[\th] + K[\ve,\eta]
    =G'[\th] + K[\ve,\eta]\,.                              \lab{2.6}
\ee
As we can see, the central term remains unchanged:
\bitem
\item[(3)]{Canonically equivalent Hamiltonian theories have
             identical central terms}.
\eitem

Going over to the Lagrangian formalism, one can make use of the above
theorem to prove two similar theorems concerning the equality of
central terms in classically equivalent Lagrangian theories. We note
that adding a pure divergence term to the Lagrangian leads to
canonically equivalent formulation of the theory. The above theorem
then implies:
\bitem
\item[(4a)]{A pure divergence term in the Lagrangian
             does not influence the central term}.
\eitem
Another way of obtaining an equivalent Lagrangian theory is to make a
different choice of variables. For example, the so-called contact
transformations $\phi\rightarrow\phi'=\phi'(\phi,t)$ are known to
produce canonically equivalent Hamiltonian theories. Indeed, since the
canonical momenta of the transformed Lagrangian
$\cL'(\phi',\pd\phi')\equiv\cL(\phi,\pd\phi)$ are
$\pi'_i\equiv\pd\cL'/\pd\dot\phi'^i = (\pd\phi^k/\pd\phi'^i)\pi_k$,
it is not difficult to verify that the transformation
$(\phi,\pi)\to(\phi',\pi')$ is canonical. As canonical transformations
do not change the central term, the following theorem is proved:
\bitem
\item[(4b)]{Contact transformations of the Lagrangian variables
             induce only trivial changes of the central term}.
\eitem
In particular, if we make the change of variables
$\phi=\bar\phi+\vphi$, where $\bar\phi$ is a fixed field configuration
(usually a solution of the field equations), the related change of the
central term is trivial.

At the end of this section, let us mention one more way of changing
Lagrangian without changing the classical field equations: multiplying
the Lagrangian by a constant, $\cL\rightarrow \a\,\cL$. As we shall see
in the next section, the central term is thereby changed as
$K\rightarrow \a\,K$. Since  the corresponding quantum theories are not
equivalent (the functional integrals defined by $\cL$ and $\a\cL$ are
different), we can say that classical central charges are not purely
classical objects after all.

\section{Central charge in the Lagrangian formalism}

In this section we express the canonical central term through the
Lagrangian coefficients, and establish a basic relation between the
canonical symmetry generator and the Noether charge.

\subsection{Canonical central charge}\label{ccc}

We consider a field theory defined by the action $I=\int d^n
x\cL(\phi,\pd\phi)$. The action is invariant (up to a boundary term)
under the symmetry transformations
\be
        \d_\ve\phi^i=R^i(\ve,\pd\ve,\phi,\pd\phi)\, ,      \lab{3.1}
\ee
which constitute a group $U$ with parameters $\ve^a$. Moreover, we
assume that the adopted asymptotic conditions are invariant under a
subgroup $U_0$ of $U$, which essentially coincides with the
asymptotic symmetry group.

After introducing the notation $\vphi^i=\phi^i-\bphi^i$, where
$\bphi$ is a fixed configuration of fields, the Lagrangian function
can be expressed as a series in $\vphi$ and $\pd\vphi$:
\bea
\cL=&&V+V_i\vphi^i +V_i^\m\pd_\m\vphi^i \nn\\
    &&+V_{ij}\vphi^i\vphi^j
      +V_{ij}^\m\pd_\m\vphi^i\vphi^j
      +V_{ij}^{\m\n}\pd_\m\vphi^i\pd_\n\vphi^j+\cO_3\, ,   \lab{3.2}
\eea
where $\cO_n$ denotes a term of the $n$-th power in $\vphi$ and/or
$\pd\vphi$, and the coefficients $(V,V_i,\dots)$ are functions of
$\bphi$. Let us now introduce the generalized momentum variables
\bea
&&\Pi_i^\m=\frac{\pd\cL}{\pd\pd_\m\phi^i}=V_i^\m
 +V_{ij}^\m\vphi^j+2V_{ij}^{\m\n}\pd_\n\vphi^j + \cO_2\, , \nn\\
&&\pi_i^\m=\Pi_i^\m-\bar\Pi_i^\m
      = V_{ij}^\m\vphi^j+2V_{ij}^{\m\n}\pd_\n\vphi^j+\cO_2\, , \nn
\eea
where $\bar\Pi^\m_i\equiv\Pi^\m_i(\phi=\bar\phi)$. The variables
$(\vphi^i,\pi_i\equiv\pi^0_i)$ are canonical, as they are obtained from
the original variables $(\phi^i,\Pi_i\equiv\Pi^0_i)$ by the canonical
transformation, but $\pi_i\ne\pd\cL/\pd\dot\vphi^i$. The action of the
symmetry transformations \eq{3.1} on $(\vphi^i,\pi_i)$ is given by
\bsubeq
\bea &&\d_\ve\vphi^i\equiv
\bR^i(\ve)+\cO_1 \,,\nn\\ &&\d_\ve\pi_i= V_{ij}^0\bR^j(\ve)
         + 2V_{ij}^{0\n}\pd_\n\bR^j(\ve)+\cO_1\, ,         \lab{3.3a}
\eea
where $\bR^i(\ve)$ is short for $R^i(\ve,\pd\ve,\bphi,\pd\bar\phi)$.

We now wish to construct the canonical generator of the transformations
\eq{3.3a}. To this end, note that both equations \eq{3.3a} and those
which define $\pi_i$ can be compared with the corresponding canonical
relations only on shell, using the Hamiltonian dynamical equations for
$\dot\vphi^i$. This is, in fact, true in general: the Lagrangian and
Hamiltonian approaches can be compared only on shell.  However, for
arbitrary $\bar\phi\,$ the Lagrangian field equations take the form
$V^i-\pd^\m V^i_\m + \cO_1(\vphi,\pd\vphi)=0$, so that, on shell,
$\cO_0$ and $\cO_1$ terms are not independent. To avoid this
conflicting situation, we assume that $\bar\phi\,$ is a {\it solution
of the field equations\/}, whereby the clean separation between $\cO_0$
and $\cO_1$ terms is maintained on shell, too. Then, the canonical
generator of the transformations \eq{3.3a} can be written in the form
\bea
G[\ve]&=&\int d^{n-1} x\,\tG(\ve)\, ,\nn\\
\tG(\ve)&=&\bR^i(\ve)\pi_i -\vphi^i\left[ V_{ij}^0\bR^j(\ve)
 +2V_{ij}^{0\n}\pd_\n\bR^j(\ve)\right]+\cO_2(\vphi,\pi)\, ,\lab{3.3b}
\eea
\esubeq
Indeed, the canonical analogs of the
transformation laws \eq{3.3a}, derived from \eq{3.3b}, contain
$\cO_1(\vphi,\pi)$ terms. Since $\pi$ is linear in
$(\vphi,\pd\vphi)$, we have
$\cO_1(\vphi,\pi)\approx\cO_1(\vphi,\pd\vphi)$ (on shell), and
consequently, $\eq{3.3a}$ follows consistently from $\eq{3.3b}$.

We see that the canonical generator \eq{3.3b} satisfies the needed
differentiability condition in the lowest order, including only
$\cO_1(\vphi,\pi)$ terms. With this accuracy, the canonical algebra
\eq{2.3} takes the form
\bsubeq\lab{3.4}
\be
\{G[\ve],G[\eta]\}= K[\ve,\eta]+\cO_1(\vphi,\pi) \, ,      \lab{3.4a}
\ee
where
\bea
K[\ve,\eta]&=&\int d^{n-1} x\,\tK(\ve,\eta)\, ,\nn\\
\tK(\ve,\eta)&=& \bR^i(\ve)\left[ V_{ij}^0\bR^j(\eta)
  +2V_{ij}^{0\n}\pd_\n\bR^j(\eta)\right]-(\ve\lra\eta)     \lab{3.4b}
\eea
\esubeq
is the classical central charge. It is identified as the zero order
term on the right-hand side of \eq{3.4a}.  The identification is unique
since $\bar\phi\,$ is a solution of the field equations.

There is an ambiguity in the derivation of the formula \eq{3.4}, which
stems from the fact that the coefficients in the $\cO_1(\vphi,\pi)$
part of the canonical generator \eq{3.3b} are not defined uniquely.
Indeed, one can change any of these coefficients by adding a term
proportional to an undetermined Hamiltonian multiplier $v(t)$ (suitably
restricted by the adopted asymptotic conditions), since such a term is
of the $\cO_0(\vphi,\pi)$ type, while its Lagrangian counterpart is
some undetermined velocity, thus of the $\cO_1$ type. Therefore, the
action of such a modified canonical generator will produce the same
$\cO_0$ terms as in \eq{3.3a}. Can one resolve this ambiguity without
examining higher order terms in the transformation law?

Let us recall that our choice of the asymptotic conditions includes the
requirement of the conservation of energy. This means that the improved
total Hamiltonian $H_T$, which contains an additive surface term
$S$, should satisfy
$$
\frac{dH_T}{d t}=\frac{\pd H_T}{\pd t}
                       =C_{PFC}+\frac{\pd S}{\pd t} \approx 0\, .
$$
Thus, $\pd S/\pd t\approx 0$, i.e. the surface integral $S$ has no
explicit time dependence, and therefore, no $v(t)$ terms. Using this
property of $H_T$ in the derivation of equation \eq{2.4}, one
can show that there are no surface terms on the right-hand side of that
equation, and consequently, all the symmetry generators are also
conserved, $d G/d t\approx 0$. Now, combining \eq{1.1} and \eq{2.4} one
finds that the central charge is time independent:
$$
\frac{\pd}{\pd t}K[\ve,\eta]=0\,,
$$
for every $\ve$ and $\eta$, and independently, for every choice of
$v(t)$. This implies that there can be no $v(t)$ terms in the central
charge $K[\ve,\eta]$. Consequently, our formula \eq{3.4} gives a
correct Lagrangian expression for the canonically defined classical
central charge.

The general theorem (4b) of the preceding section states that the
choice of $\bphi$ is inessential for the value of the central charge,
as a consequence of the canonical invariance of the theory. On the
other hand, one can see from our result \eq{3.4} that $K[\ve,\eta]$
explicitly depends on $\bphi$. This apparent contradiction is
resolved by noting that our choice of $\bphi$ is naturally adjusted
to the asymptotic properties of the theory, in the sense that
\bitem
\item[$(a)$] $\bphi$ is a solution of the field equations, and
\vspace{-10pt}
\item[$(b)$] it satisfies the adopted asymptotic conditions.
\eitem
As a consequence, the original {\it canonical invariance is broken\/},
and $\bphi$ cannot be changed arbitrarily any more. Still, there
remains the freedom to replace $\bphi$ by another field configuration
satisfying $(a)$ and $(b)$, which produces only a trivial modification
of the central charge. The dependence of $K[\ve,\eta]$ on $\bphi$ is
quite natural, as our choice of $\bphi$ embodies the basic structure of
the asymptotic symmetry.

For further analysis and comparison with the Lagrangian formalism,
it is useful to define the ``canonical current"
\bsubeq\lab{3.5}
\be
\tG^\m(\ve)=\vphi^i\left[ V_{ij}^\m\bR^j(\ve)
          +2V_{ij}^{\m\n}\pd_\n\bR^j(\ve)\right]
      -\bR^i(\ve)\left[ V_{ij}^\m\vphi^j
          +2V_{ij}^{\m\n}\pd_\n\vphi^j\right]+\cO_2\, ,    \lab{3.5a}
\ee
as a generalization of the canonical generator density $\cal G$
(up to a sign), and the ``central current"
\be
\tK^\m(\ve,\eta)\equiv\bR^i(\ve)\left[ V_{ij}^\m\bR^j(\eta)
  +2V_{ij}^{\m\n}\pd_\n\bR^j(\eta)\right]-(\ve\lra\eta)\,,\lab{3.5b}
\ee
\esubeq
as a generalization of the central charge density $\cal K$. These
two currents satisfy the relation
\bea
\d_\ve\tG^\m(\eta)=\tK^\m(\ve,\eta)+\cO_1\, ,              \lab{3.6}
\eea
which is the ``current" version of the canonical algebra \eq{3.4a}.

\subsection{Noether current}\label{NC}

The symmetry transformation \eq{3.1} leaves the action invariant, up
to a boundary term: $\d_\ve I=\int_{\cM} dK$, where
$K=K(\ve,\pd\ve,\phi,\pd\phi)$ is an $(n-1)$-form on
$\cM$. Since $dK=\pd_\m k^\m d^n x$, where
$k^\m\equiv \ve^{\m\m_1\dots\m_{n-1}}K_{\m_1\dots\m_{n-1}}/(n-1)!$,
this identity can be written as
\bsubeq\lab{3.7}
\be
\d_\ve\cL=\pd_\m k^\m\, .                                  \lab{3.7a}
\ee
The coefficients $k^\m$ are the components of the vector-valued $n$-form
$k^\m\pd_\m\otimes d^n x$. Equation \eq{3.7a} implies
\be
F_i\d_\ve\phi^i =\pd_\m N^\m\, .                           \lab{3.7b}
\ee
Here, $F_i=\d\cL/\d\phi^i$ are the field equations, and $N^\m$ is
the (conserved) Noether current:
\be
N^\m = k^\m-\frac{\pd\cL}{\pd\pd_\m\phi^i}\d_\ve\phi^i
     \equiv N^\m(\ve)   \, ,                               \lab{3.7c}
\ee
\esubeq
where, as usual, $N^\m(\ve)$ is short for
$N^\m(\ve,\pd\ve,\phi,\pd\phi)$. When $U_0$ is a Lie group, the
Noether current takes the form $N^\m(\ve)=\ve_a
N^{a\m}(\phi,\pd\phi)$. One should note that $N^\m$ is defined only
up to a trivial term $\pd_\r N^{\m\r}$, with $N^{\m\r}=-N^{\r\m}$. We
shall show that there is a suitable choice of the superpotential
$N^{\r\m}$, such that $\tG^\m=N^\m+\cO_2$, which identifies $N^\m$ as
the Lagrangian equivalent of $\tG^\m$.

As $\bphi$ is a solution of the field
equations, it holds $V_i=\pd_\m V_i^\m$, which implies
\bea
&&\cL=V+\pd_\m(V_i^\m\vphi^i) +\cL_2\nn\\
&&\cL_2\equiv V_{ij}\vphi^i\vphi^j+V_{ij}^\m\pd_\m\vphi^i\vphi^j
      +V_{ij}^{\m\n}\pd_\m\vphi^i\pd_\n\vphi^\n+\cO_3\, .\nn
\eea
The current $k^\m$ is now given by
$$
k^\m=V_j^\m R^j+\ell^\m\, ,\qquad\d_{\ve}\cL_2=\pd_\m\ell^\m\, .
$$
Looking for $\ell^\m$ in the form
$$
\ell^\m=\a^\m+\a^\m_j\vphi^j+\a_j^{\m\n}\pd_\n\vphi^j+\cO_2\, ,
$$
and using the arbitrariness of $\vphi$ and $\pd\vphi$, we find that
the coefficients satisfy the following set of cascade equations:
\bea
&&\pd_\m\a^\m=0\, ,                                  \nn\\
&&\pd_\m\a_j^\m=2V_{ij}\bR^i+V^\m_{ij}\pd_\m\bR^i\, ,\nn\\
&&\a_j^\m+\pd_\r\a_j^{\r\m}=
     V_{ji}^\m\bR^i +2V_{ji}^{\m\r}\pd_\r\bR^i\, ,   \nn\\
&&\a_j^{\m\r}+\a_j^{\r\m}=0\, .                            \lab{3.8}
\eea
The first equation implies $\a^\m=\pd_\r\a^{\m\r}$, with
$\a^{\m\r}=-\a^{\r\m}$, so that
\bea
&&\ell^\m=(\a^\m_j-\pd_\r\a_j^{\m\r})\vphi^j
          +\pd_\r\ell^{\m\r}+\cO_2\, ,   \nn\\
&&\ell^{\m\r}\equiv(\a^{\m\r}+\a_j^{\m\r}\vphi^j) \, . \nn
\eea
Disregarding the trivial term $\pd_\r\ell^{\m\r}$ leads to
$$
k^\m= V_j^\m R^j +\left(V_{ji}^\m\bR^i
     +2V_{ji}^{\m\r}\pd_\r\bR^i\right)\vphi^j+\cO_2\, .
$$
Finally, combining this result with the definition \eq{3.7c} of the
Noether current, we obtain
\bea
N^\m&=& k^\m-R^i(V_i^\m+V_{ij}^\m\vphi^j
                 +2V_{ij}^{\m\n}\pd_\n\vphi^j)+\cO_2         \nn\\
 &=&\vphi^j\left(V_{ji}^\m\bR^i +2V_{ji}^{\m\r}\pd_\r\bR^i\right)
   -\bR^i(V_{ij}^\m\vphi^j+2V_{ij}^{\m\n}\pd_\n\vphi^j)+\cO_2 \nn\\
   &=&\tG^\m+\cO_2\, .                                     \lab{3.9}
\eea
Let us stress that this equality holds only if we choose
$\pd_\r\ell^{\m\r}=0$. Accordingly, equation \eq{3.9} can be understood
as a condition which defines the choice of the superpotential
$N^{\m\r}$.

The conditions \eq{3.8} on $\a_i^\m$ imply the identity
\be
2V_{ij}\bR^i+V^\m_{ij}\pd_\m\bR^i=
\pd_\m\left( V_{ji}^\m\bR^i
            +2V_{ji}^{\m\r}\pd_\r\bR^i\right)\, ,          \lab{3.10}
\ee
directly related to the invariance condition \eq{3.7}. Using this
identity, we easily derive the relation
$$
\pd_\m\tK^\m(\ve,\eta)=0\, ,
$$
which implies $\tK^\m=\pd_\r\tK^{\m\r}$, with $\tK^{\m\r}=-\tK^{\r\m}$
(Poincar\'e lemma). Consequently, the central term can be represented
as a surface integral:
\be
K[\ve,\eta]=\int d^{n-1}x\,\tK^0(\ve,\eta)
           =\int dS_\b\tK^{0\b}(\ve,\eta) \, .             \lab{3.11}
\ee
The conditions of the Poincar\'e lemma ensure that this result holds
true in topologically trivial spaces, while in general, central charges
may fail to be represented by surface integrals.

\subsection{Algebra of charges}

As we have seen in the preceding subsection, the Noether current $N^\m$
can be chosen to equal the canonical current ${\cal G}^\m$ in the
lowest order. This allows us to rewrite \eq{3.6} in a pure Lagrangian
form:
\bea
\d_\ve N^{\m}(\eta)=\tK^{\m}(\ve,\eta)+\cO_1 \,.           \lab{3.12}
\eea
It seems that this relation can be used as the Lagrangian definition of
the central charge. Here, the Noether current is identified with the
canonical current $\tG^\m$. Can we similarly identify the Noether
conserved charges $Q[\ve]$ with the canonical generators $G[\ve]$? A
short analysis shows that the answer is affirmative. Namely, it is well
known that the well defined canonical generators $G[\ve]$ are also
conserved quantities of the theory. If we define the Lagrangian charges
\be
Q\equiv G\bigg[\phi,\Pi=\frac{\pd\cL}{\pd\dot\phi}\bigg]
             _{\phi=\bar\phi+\vphi} \, ,                   \lab{3.13}
\ee
they will certainly be conserved as the on-shell values of the
conserved generators. Moreover, the transformation properties of the
charges $Q[\ve]$ will match those of $G[\ve]$:
\bsubeq\lab{3.14}
\be
\d_{\eta}Q[\ve]\approx Q[\th]+K[\ve,\eta] \, .             \lab{3.14a}
\ee
This is a simple consequence of the fact that symmetry
transformations do not change the field equations. The weak equality
in \eq{3.14a} means the equality up to terms proportional to the
equations of motion. Notice, however, there is no physical
distinction between $Q[\ve]$ and $Q[\ve]+\cO(F)$ since their
on-shell values are the same. It is, therefore, natural to work with
the equivalence classes of charges $\langle Q\rangle$, defined by
\[
\langle Q_1\rangle=\langle Q_2\rangle \quad \Leftrightarrow
  \quad Q_1 \approx Q_2 \,.
\]
In terms of the equivalence classes, the algebra of charges is given
by the strong equality
\be
\d_{\eta} \langle Q[\ve]\rangle=\langle Q[\th]\rangle
                                        +K[\ve,\eta] \, .  \lab{3.14b}
\ee
\esubeq
The algebra of charges \eq{3.14} yields a Lagrangian definition of the
central charge. Notice, however, that neither the Noether current
$N^\m(\ve)$ nor the charge $Q[\ve]$ are uniquely defined. The closure
of the algebra \eq{3.14} is obtained only if the existing arbitrariness
is fixed in a definite, precise way. We have seen that this can be
achieved by identifying Lagrangian charges with the on-shell values of
canonical generators, and choosing the asymptotic conditions which make
canonical generators well defined and ensure the conservation of
energy. If we do not want to refer to the Hamiltonian theory, the only
way to find the proper Noether superpotential and boundary conditions
is to demand finiteness and conservation of Noether charges \cite{14}.

\section{Central charge in gauge theories}

Gauge theories are by far the most interesting field theories from
the physical point of view. In the case of gauge symmetries, we
derive an explicit representation of the central term as a surface
integral.

\subsection{Noether identities}

Consider a gauge theory with the transformation law
\be
\d_\ve \phi^i=R^i_a\ve^a+R^{i\m}_a\pd_\m\ve^a\, .          \lab{4.1}
\ee
In this case, the invariance condition \eq{3.7b} can be written in
the form
\be
\left[F_i R^i_a-\pd_\m(F_i R^{i\m}_a)\right]\ve^a=\pd_\m J^\m\, ,
\qquad J^\m\equiv N^\m-F_iR^{i\m}_a\ve^a\, .               \lab{4.2}
\ee
Let $\Om$ be a region in spacetime, and let the gauge parameters
$\ve^a=\ve^a(x)$ satisfy the conditions: a) $\ve^a$ are arbitrary in
$\Om$, and b) $\ve^a$ and their derivatives vanish on $\pd\Om$.
Then, the integration of equation \eq{4.2} over $\Om$ leads
to the following identities:
\bsubeq
\bea
&& F_i R^i_a-\pd_\m(F_i R^{i\m}_a)=0\, ,                 \lab{4.3a}\\
&& \pd_\m J^\m=0 \, .                                      \lab{4.3b}
\eea
The first identity is known as the Noether identity, and the second
one implies (Poincar\'e lemma) that
\be
J^\m=\pd_\r N^{\m\r}\, ,\qquad N^{\m\r}=-N^{\r\m}\, .      \lab{4.3c}
\ee
\esubeq
As a consequence, we find that Noether current has the form
\be
N^\m=F_iR^{i\m}_a\ve^a+\pd_\r N^{\m\r}\, .                 \lab{4.4}
\ee
One should stress that, here, $N^{\m\r}$ is a completly arbitrary
superpotential. Its value is fixed by the extra condition \eq{3.9},
which relates Noether and canonical structures of the theory.

Having in mind our goal, to find the Lagrangian structure of the
canonical central term,  it is useful to expand the Noether identity
in terms of $\vphi$ and $\pd\vphi$. Starting from the Lagrangian
\eq{3.2} where $\phi=\bphi$ is a solution of the field equations,
$V_i-\pd_\m V_i^\m=0$, we obtain
\bea
F_i&&=(2V_{ij}-\pd_\r V_{ij}^\r)\vphi^j
      +2(V_{ji}^\n-\pd_\r V^{\n\r}_{ji})\pd_\n\vphi^j
      -2V_{ij}^{\m\n}\pd_\m\pd_\n\vphi^j+\cO_2  \nn\\
   &&\equiv f_{ij}\vphi^j+f_{ij}^\n\pd_\n\vphi^j
            +f_{ij}^{\m\n}\pd_\m\pd_\n\vphi^j+\cO_2\, .    \lab{4.5}
\eea
Using this result, the Noether identity \eq{4.3a} takes the form
given in Appendix A, which will be usefull in further considerations.

\subsection{Central term as a surface integral}

Starting from the equality  $N^\m=\tG^\m+\cO_2$, equation \eq{3.9},
it follows that $\tG^\m$ has the Noether form \eq{4.4}:
\bsubeq\lab{4.6}
\be
\tG^\m(\ve)=F_i\bR^{i\m}_a\ve^a +\pd_\r N^{\m\r}+\cO_2\, . \lab{4.6a}
\ee
Using the expression \eq{3.5a} for $\tG^\m(\ve)$, one finds that
$N^{\m\l}$ is given by (Appendix B):
\bea
N^{\m\l}=&&\bigg\{ \left[(V_{ji}^\m-\pd_\r V_{ji}^{\m\r})\vphi^j
  -2V_{ji}^{(\m\r)}\pd_\r\vphi^j \right]\bR^{i\l}_a\ve^a   \nn\\
  && + V_{ji}^{\m\l}\vphi^j\bR^i
  +\frac{2}{3}\pd_\r(V_{ji}^{\m\r}\vphi^j\bR^{i\l}_a\ve^a)\bigg\}
    -(\m\lra\l) +\cO_2\, .                                 \lab{4.6b}
\eea
\esubeq
Now, the relation $\d_\ve\tG^\m(\eta)=\tK^\m(\ve,\eta)+\cO_1$ and the
above expression for $\tG^\m$ imply
\bsubeq\lab{4.7}
\be
\tK^\m(\ve,\eta)=\pd_\l\left[\d_\ve N^{\m\l}(\eta)
                 \right]_{\vphi=0} \, .                    \lab{4.7a}
\ee
Consequently, the central term is expressed as the following
surface integral:
\be
K[\ve,\eta]=\int d^{n-1}x\,\tK^0(\ve,\eta)=\int dS_\b
\left[\d_\ve N^{0\b}(\eta)\right]_{\vphi=0}\,.             \lab{4.7b}
\ee
\esubeq

\section{Examples}

\subsection{Chern-Simons theory}

As our first example, we consider a Chern-Simons gauge theory on a 3d
spacetime $\cM$ with the topology $\cM=R\times\S$, where $R$ is
interpreted as time, and $\S$ is a spatial manifold whose boundary is
topologically a circle located at infinity. For the gauge group
$SL(2,R)$, the theory is defined by the Lagrangian
\bsubeq
\be
\cL(A,\pd A)=\k\ve^{\m\n\r}\left(A^a{_\m}\pd_\n A_{a\r}
    +\frac{1}{3}\ve_{abc}A^a{_\m}A^b{_\n}A^c{_\r}\right)\, .\lab{5.1}
\ee
The action is invariant under the infinitesimal gauge transformations
\be
\d_\t A{_\m}=\nabla_\m\t\equiv \pd_\m\t+[A{_\m},\t] \, ,   \lab{5.1b}
\ee
where $A_\m= A^a{_\m}T_a$, $\t=\t^aT_a$, and $T_a$ is the basis of
the Lie algebra $s\ell(2,R)$, defined by $[T_a,T_b]=\ve_{ab}{^c}T_c$.
The field equations have the form
\be
F_{\m\n}=\pd_\m A_\n-\pd_\n A_\m+[A_\m,A_\n]=0\, .         \lab{5.1c}
\ee
\esubeq
Every solution of the field equations is a pure gauge:
$A_\m=g^{-1}\pd_\m g$, where $g$ is in $SL(2,R)$.

We choose the asymptotic conditions which, in polar coordinates
$x^\m=(t,r,\vphi)$, read:
\be
A_+\equiv\frac{1}{2}( A_0+A_2)=\cO_1\, ,                   \lab{5.2}
\ee
and $\cO_1$ is a term that tends to zero as $1/r$ or faster when
$r\to\infty$. The conditions \eq{5.2} imply that gauge parameters are
independent of $x^+=x^0+x^2$ in the asymptotic region. In what follows,
we shall often refine the asymptotics by the assumption that field
equations decrease arbitrarily fast in spatial infinity, since no
solution of the field equations is thereby lost. Using this principle,
we find that $\pd_+A_{\m}=\cO_1$, whereupon the asymptotic dependence
on $x^+$ is completely eliminated from the theory.

In what follows, we shall analyze two typical situations belonging to
the class of asymptotic conditions defined above. As before, our fields
are expanded around a classical solution,
\be
A{_\m}=\bA_\m+\cA_\m \, , \qquad
    \d_\t\cA_\m=\bar\nabla_\m\t+[\cA_\m,\t]\, ,            \lab{5.3}
\ee
where $\bA_\m$ satisfies \eq{5.2}.

\prg{1.} We begin by noting that gauge transformations in \eq{5.3}
are defined in terms of the gauge parameters $\t^a$. The related
commutator algebra has the form
$[\d_\l,\d_\t]\cA^a{_\m}=\d_\th\cA^a{_\m}$, where
$\th^a=\ve^a{}_{bc}\l^b\t^c$. The central term is defined by the
relation \eq{3.4}, where
$$
\bR^i(\t) V_{ij}^0\bR^j(\l)-(\t\lra\l)
          = 2\k\ve^{0\a\b}\bar\nabla_\a\t^a\bar\nabla_\b\l_a\, .
$$
The integration of this expression over the spatial section of
spacetime leads to
\bea
K[\t,\l]&=&2\k\int_\S d^2 x\,\ve^{0\a\b}
                          \bar\nabla_\a\t^a\bar\nabla_\b\l_a \nn\\
&=& 2\k\int_{\pd\S}dS_\a\ve^{0\a\b}(\t^a\pd_\b\l_a+\th^a\bA_{a\b})\,,\nn
\eea
where we used the Stokes theorem and $\bar F^a{}_{\a\b}=0$. The
second term in this expression is a functional of $\th$, hence it
represents a trivial contribution to the central charge. Ignoring
this term, we find that the essential piece of $K$ is given by
\be
K_\ess[\t,\l]= 2\k\int dS_\a\,\ve^{0\a\b}\t^a\pd_\b\l_a
           =2\k\int_0^{2\pi}d\vphi\,\t^a\pd_\vphi\l_a \, . \lab{5.4}
\ee
The same result follows from the surface integral expression
\eq{4.7b}, as expected. At the canonical level, the related
asymptotic symmetry is described by the affine (or Kac-Moody)
extension of $s\ell(2,R)$ \cite{8,9}.

\prg{2.} The second case is defined by a set of additional asymptotic
conditions, which restrict the original asymptotic symmetry, as
described bellow.

(a) First, we replace $\t^a$ with the new gauge parameters $\xi^\m$,
according to
$$
\t^a=\xi^\m A^a{_\m}\, .
$$
The composition law is defined by
$[\d_\eta,\d_\xi]\cA^a{_\m}=\d_\chi\cA^a{_\m}$, with
$\chi^\m=\xi\cdot\pd\eta^\m -\eta\cdot\pd\xi^\m$. The new parameters
$\xi^\m$ correspond to diffeomorphisms, which is related to the
gravitational content of the Chern-Simons theory. Note that, as a
consequence of $A{_+}=\cO_1$, the parameter $\xi^+$ is effectively
absent from the asymptotic region, $\xi^\m A{_\m}=\xi^r A{_r}+\xi^-
A{_-}+\cO_1$. As before, $\pd_+A_{\m}=\cO_1$ and $\pd_+\t=\cO_1$, which
yields $\pd_+\xi^\m=\cO_1$.

(b) In the second step, the asymptotic behaviour of $A_{\m}$ fields is
further restricted by the requirement
$$
A_r=\a+\cO_1\, ,
$$
where $\a$ is a constant element of the Lie algebra. Again, the field
equations can be used to refine the asymptotics. This leads to
$$
A_-=e^{-r\a}\Om_-(x^-)e^{r\a}+\cO_1  \, ,
$$
with $\Om_-(x^-)$ an arbitrary Lie algebra valued function of $x^-$. The
new asymptotic conditions restrict the diffeomorphism parameters
$(\xi^r,\xi^-)$ to be independent of $r$, and $\t\equiv\xi^\m A_\m$ takes
the form $\t=e^{-r\a}u(x^-)e^{r\a}+\cO_1$. The conditions $A_+=\cO_1$ and
$A_r=\a +\cO_1$ are clearly invariant under the residual symmetry
transformations of this form, while $\Om_-$ is transformed as follows:
$$
\d\Om_-=\pd_-u +[\Om_-,u]\, .
$$

(c) Motivated by the relevant considerations in 3D gravity, we
restrict $(\xi^r,\xi^-)$ to be of the specific form \cite{8,9}:
$$
(\xi^r,\xi^-)=(\pd_\vphi\xi,\xi)\, .
$$
Such a choice leads to the conformal asymptotic symmetry of 3D
gravity. It is straightforward to verify that the restricted
parameters are consistent with the composition law of
diffeomorphisms.

For the classical solution $\bA$, which belongs to the set of
asymptotic states described above, we choose
$$
\bA_+=0\, ,\quad  \bA_r=\a\, ,\quad \pd_+\bar A_-=0\, .
$$
Using the change of parameters introduced in (a), one finds that the
central charge of the restricted symmetry $K[\xi,\eta]$ is obtained
from $K[\t,\l]$ by the replacements
$\t^a\to\bar\t^a\equiv\xi^\m\bA^a{_\m}$ and
$\l^a\to\bar\l^a\equiv\eta^\m\bA^a{_\m}$:
$$
K[\xi,\eta]=2\k\int_{\pd\S}dS_\a
            \ve^{0\a\b}(\bt^a\pd_\b\bl_a+\bth^a\bA_{a\b})\, .
$$
The property $\bA_+=0$ implies the vanishing of the last term, since
$\ve^{abc}\bA_{a\m}\bA_{b\n}\bA_{c\b}=0$. Then, one can derive an
important identity, which immediately leads to
\bsubeq
\bea
&&K[\xi,\eta]=K_\ess[\xi,\eta]
            +2\k\int d\vphi\left(\a\cdot\bar\Om_-\,\chi^r
            -\frac{1}{2}\bar\Om_-^2\,\chi^-\right) \, ,\nn\\
&&K_\ess[\xi,\eta]\equiv
             2\k\a^2 \int d\vphi\,\xi^r\pd_\vphi\eta^r\, . \lab{5.5a}
\eea
The second term in $K[\xi,\eta]$ is a trivial piece of the central
charge. Finally, replacing here the restricted gauge parameters as
defined in (c), we obtain
\be
K_\ess[\xi,\eta]=2\k\a^2\int_0^{2\pi} d\vphi\,
                \pd_\vphi\xi\pd^2_\vphi\eta\, .            \lab{5.5b}
\ee
\esubeq
The structure of the asymptotic symmetry obtained in this way is
described by the Virasoro algebra with classical central term,
$c=-12\times 4\pi\k\a^2$ \cite{8,9}.

If we change $\bA_-$ and keep $\bA_r=\a$ fixed, the change of the
central charge is trivial. To understand the dependence of $K_\ess$
on $\a$, it is useful to recall that $e^{r\a}$ defines the holonomy
of the flat connection $A_\m$ \cite{11}. In the canonical
framework, $\a$ defines the values of the conserved asymptotic
generators. Thus, $\a$ is an essential characteristic of the chosen
asymptotic configuration, and consequently, changing $\a$ would mean
a transition to another, non-equivalent asymptotic configuration,
possessing a different central charge.

It is interesting to note that 3d gravity can be formulated in terms of
two independent Chern-Simons gauge theories. In Riemannian or
teleparallel theory we have $4\pi\k=\ell/8G$, $\a^2=-1$, and
$c_1=c_2=3\ell/2G$, while in Riemann-Cartan gravity, one can have two
different central charges \cite{20}.

\subsection{Liouville theory}

As our second example, we consider the Liouville theory on a cylinder.
Using the coordinates $\,x^0=t\in (-\infty,+\infty)$, $\, x^1=\vphi\in
[0,2\p)$, the Lagrangian and the related equations of motion read
\be
\cL = \frac{1}{2}(\pd_{\m}\phi)(\pd^{\m}\phi)-
      \frac{\m^2}{\a^2}e^{\a\phi} \,, \qquad
\Box\phi + \frac{\m^2}{\a}e^{\a\phi}=0 \,.                 \lab{5.6}
\ee
The transformation law
\be
\d_{\ve}\phi = -\ve^{\m}\pd_{\m}\phi
               -\frac{1}{\a}\pd_{\m}\ve^{\m}               \lab{5.7}
\ee
defines the symmetry of the theory, provided the parameters $\ve^{\m}$
satisfy the conditions
$$
\pd_\mp \ve^\pm=0\quad\Ra\quad \ve^\pm=\ve^\pm(x^\pm)\, ,
$$
where $x^\pm=x^0\pm x^1$, and similarly for $\ve^\pm$.

Let us now define the variable $\vphi\equiv \phi - \bphi$, where
$\bphi$ is a fixed field configuration. Then, the Lagrangian takes the
polynomial form \eq{3.2} with the coefficients $V_i^{\m}\equiv V^{\m}$
and $V_i^{\m\n}\equiv V^{\m\n}$ given by
\be
V^{\m}=0\,,\qquad V^{\m\n}=\frac{1}{2}\,\eta^{\m\n}\, .    \lab{5.8}
\ee
At the same time, the transformation law becomes
\be
\d_{\ve}\vphi = \bR(\ve)-\ve^{\m}\pd_{\m}\vphi\,, \qquad
\bR(\ve)\equiv -\ve^{\m}\pd_{\m}\bphi
               -\frac{1}{\a}\pd_{\m}\ve^{\m} \,.           \lab{5.9}
\ee
The composition law of the group is obtained from the commutator
algebra of these transformations:
$[\d_{\eta},\d_{\ve}]\vphi =\bR(\th)-\th^{\m}\pd_{\m}\vphi$, where
$\th^{\m}\equiv \eta^{\n}\pd_{\n}\ve^{\m}-\ve^{\n}\pd_{\n}\eta^{\m}$.

Now, we are ready to calculate the central charge. Using the formula
\eq{3.5b}, with the coefficients \eq{5.8}, we find
$$
\cK^\m(\ve,\eta) = \eta^{\m\n}\bR(\ve)\pd_\n\bR(\eta)-(\ve\lra\eta)\,,
$$
wherefrom
\bea
\cK^0(\ve,\eta) =
&&\bigg\{
\frac{1}{\a^2}\left[(\pd_{+}\ve^{+})(\pd^2_{+}\eta^{+}) +
(\pd_{-}\ve^{-})(\pd^2_{-}\eta^{-})\right]                     \nn\\
&&+(\ve^{+}\eta^{-})(\Box\bphi)\,\pd_1\bphi +
\frac{1}{\a}(\Box\bphi)\,\pd_1(\ve^{+}\eta^{-})\bigg\} -
(\ve \leftrightarrow \eta)+\pd_1 D+\cO(\th) \,.            \lab{5.10}
\eea
In the above formula, the terms $\pd_1 D$ and $\cO(\th)$ are not
explicitly displayed because they do not influence the central charge. The
first is a spatial derivative of some quantity $D$, and consequently,
vanishes after the integration over the compact spatial section of the
cylinder. The second gives a trivial contribution to the central term
since its dependence on $\ve$ and $\eta$ is only through $\th(\ve,\eta)$.

We continue the analysis of the central charge by recalling that
$\bphi$ must be a solution of the field equations. Using this fact,
we find that the whole $\bphi$ dependent part of the equation \eq{5.10}
becomes a spatial derivative, and can be included in $\pd_1 D$. The
final expression for the central charge is obtained by integrating
$\cK^0(\ve,\eta)$ over the spatial section of the cylinder:
\be
K_\ess[\ve,\eta] = \frac{1}{\a^2}\int_0^{2\p}d\vphi
              \left[(\pd_{+}\ve^{+})(\pd^2_{+}\eta^{+}) +
              (\pd_{-}\ve^{-})(\pd^2_{-}\eta^{-})\right] -
              (\ve \leftrightarrow \eta) \, .              \lab{5.11}
\ee

Two comments are in order: a) The obtained expression for central
charge has no explicit $\bphi$ dependence. This is a consequence of the
fact that $\bphi$ satisfies both conditions $(a)$ and $(b)$ given
in section \ref{ccc}. b) The central charge \eq{5.11} is nontrivial
despite the fact that the spatial section of the cylinder has no
boundary. This follows from our observation in section \ref{NC}, that
central terms are not always given as boundary integrals.

\subsection{General relativity}

In the tetrad formalism, general relativity with a cosmological
constant is described by the Lagrangian density
\bsubeq
\bea
\cL&=&-abR(A)-2a\L b \nn\\
   &=&\frac{1}{2}\,a\,\ve^{\m\n\l\r}_{ijkl}b^k{_\l}b^l{_\r}
      \left(\pd_\m A^{ij}{_\n}+A^i{}_{s\m}A^{sj}{}_\n\right)
      -2a\L b \, ,                                         \lab{5.12a}
\eea
where $b^i{_\m}$ are tetrad fields, and $A^{ij}{_\m}$ is the spin
connection. Gauge symmetries of the theory are local translations and
local Lorentz rotation, with parameters $\ve^a=(\ve^\m,\ve^{ij})$:
\bea
&&\d_\ve b^i{_\m}=\ve^i{_s}b^s{_\m}-(\pd_\m\ve^\l)b^i{_\l}
                  -\ve^\r\pd_\r b^i{_\m}\, ,              \nn\\
&&\d_\ve A^{ij}{_\m}=-\nabla_\m\ve^{ij}-(\pd_\m\ve^\l)A^{ij}{_\l}
                  -\ve^\r\pd_\r A^{ij}{}_\m\, .            \lab{5.12b}
\eea
\esubeq
The commutator algebra $[\d_\ve,\d_\t]b^i{_\m}=\d_\th b^i{_\m}$ defines
the composition law:
\bea
&&\th^i{_k}=\,\t^i{_m}\ve^m{_k}-\t\cdot\pd\ve^i{_k}-(\ve\lra\t)\, ,\nn\\
&&\th^\r=\,\t\cdot\pd\ve^\r -\ve\cdot\pd\t^\r \, . \nn
\eea

After introducing
$$
b^i{_\m}=\bb^i{_\m}+\cB^i{_\m} \, ,\qquad
A^{ij}{_\m}=\bA^{ij}{_\m}+\cA^{ij}{_\m} \, ,
$$
where $(\bar b^i{_\m},\bar A^{ij}{_\n})$ is a solution of the field
equations, we can expand $\cL$ in powers of $(\cB,\cA)$, leading to
\be
V_{ij}^{\m\n}=0 \, ,\qquad
V_{ij}^\m(\pd_\m\vphi^i)\vphi^j=\ve^{\m\n\l\r}_{ijkl}
         \bb^k{_\l}\cB^l{_\r}\pd_\m\cA^{ij}{_\n}\, .       \lab{5.13}
\ee
Then, equation \eq{4.7b} yields directly the surface integral
expression for the central charge. Indeed, we have
\bea
N^{0\b}(\t)&=& V_{ij}^0\vphi^i\bR^{j\b}_a\t^a
        -V_{ij}^\b\vphi^i\bR^{j 0}_a\t^a       \nn\\
       &=& 2a\ve^{0\a\g\b}_{ijkl}\bb^k{_\g}
            (-\t^\r\bb^l{_\r})\cA^{ij}{_\a}\, , \nn
\eea
and consequently,
\bea
K[\ve,\t]&=&
       \int dS_\b\left[\d_\ve N^{0\b}(\t)\right]_{\cA,\,\cB=0} \nn\\
         &=&2a\int dS_\b\, \ve^{0\a\g\b}_{ijkl}\bb^k{_\g}\bb^l{_\r}
            \left[\bar\nabla_\a\ve^{ij}+(\pd_\a\ve^\l)\bA^{ij}{_\l}
           +\ve\cdot\pd\bA^{ij}{_\a} \right]\t^\r\, .      \lab{5.14}
\eea
This formula is the tetrad counterpart of the metric expression for
central charge in general relativity \cite{1,15}.

\section{Concluding remarks}

In the canonical formalism, asymptotic symmetry is characterized by the
PB algebra, which may  have central extension. Motivated by this
canonical structure, we carried out an investigation of the central
term from the Lagrangian point of view. Our analysis is based on the
assumption that the adopted asymptotic configuration of fields
guarantees that the canonical generators are finite, differentiable and
conserved. A simple canonical derivation of the central term leads to
the basic formula \eq{3.4}. This result is then transformed into the
Lagrangian definition of the central term \eq{3.14}, where the role of
canonical charges is taken over by suitable normalized Noether charges.
In the case of gauge theories, we found the general surface integral
expression \eq{4.7} for the central term. The results are then applied
to several interesting examples, which are intended to illustrate the
method. We expect that the present analysis of central term could be
extended to a complete Lagrangian treatment of asymptotic symmetries,
fully equivalent with the canonical structure \eq{1.1}.

\acknowledgements
This work was supported by the Serbian Science Foundation, Serbia.

\appendix

\section{Different forms of the Noether identity}

In this appendix, we derive two alternative forms of the Noether
identity.

After using the expansion \eq{4.5} for $F_i$ in \eq{4.3a}, the
condition that the coefficients multiplying fields $\vphi^i$ and
their derivatives vanish, leads to
\bea
&&\bR^i_a f_{ij}=\pd_\r(\bR^{i\r}_a f_{ij})+\cO_1\, ,       \nn\\
&&\bR^i_a f_{ij}^\l= \bR^{i\l}_a f_{ij}
               + \pd_\r(\bR^{i\r}_a f^\l_{ij})+\cO_1\, ,    \nn\\
&&\bR^i_a f_{ij}^{(\m\l)}= \bR^{i(\m}_a f^{\l)}_{ij}
               +\pd_\r(\bR^{i\r}_a f^{(\m\l)}_{ij})+\cO_1\,,\nn\\
&&0=\sum_{\pi(\m\l\r)}\bR^{i\m}_a f^{\l\r}_{ij}\, ,        \lab{A.1}
\eea
where the sum goes over all permutations $\pi$ of $(\m,\l,\r)$ (the
symmetrization).

The identity \eq{3.10} is obtained from the invariance condition
\eq{3.7a}. We now show that for gauge theories, this relation reduces
to the form \eq{A.1}. Indeed, replacing the expression \eq{4.1} for
$\bR^i$ into \eq{3.10}, the arbitrariness of $\ve^a$ and $\pd\ve^a$
leads to the relations
\bea
V_{ij}^\r\pd_\r \bR^i_a+2V_{ij}\bR^i_a
                  &=&\pd_\m(V_{ji}^\m\bR^i_a)
     + 2\pd_\m(V_{ji}^{\m\r}\pd_\r\bR^i_a) \, ,           \nn\\
V_{ij}^\m\bR^i_a+V_{ij}^\l\pd_\l\bR^{i\m}_a
     +2V_{ij}\bR^{i\m}_a&=& V_{ji}^\m\bR^i_a
     + \pd_\l(V_{ji}^\l\bR^{i\m}_a)                       \nn\\
  && + 2V_{ji}^{\m\r}\pd_\r\bR^i_a
     + 2\pd_\l(V_{ji}^{\l\m}\bR^i_a)
     + 2\pd_\l(V_{ji}^{\l\r}\pd_\r\bR^{i\m}_a) \, ,       \nn\\
2V_{ij}^{(\r}\bR^{i\m)}_a&=&  2V_{ji}^{(\m\r)}\bR^i_a
     +2V_{ji}^{(\m\l}\pd_\l\bR^{i\r)}_a
     +2\pd_\l(V_{ji}^{\l(\r}\bR^{i\m)}_a)\, ,             \nn\\
 0&=&\sum_{\pi(\r\m\l)} V_{ji}^{\r\m}\bR^{j\l}_a \, ,     \nn
\eea
which can be rewritten in an equivalent form as
\bea
\bR^i_a f_{ij} &=&2\pd_\m\left(V_{ji}^\m\bR^i_a
     + V_{ji}^{\m\r}\pd_\r\bR^i_a\right) \, ,             \nn\\
\bR^{i\m}_af_{ij}&=& 2\left( V_{ji}^\m\bR^i_a
     + V_{ji}^{\m\r}\pd_\r\bR^i_a\right)                  \nn\\
  && +2\pd_\l\left( V_{ji}^\l\bR^{i\m}_a
     + V_{ji}^{\l\m}\bR^i_a
     + V_{ji}^{\l\r}\pd_\r\bR^{i\m}_a\right) \, ,         \nn\\
V_{ij}^{(\r}\bR^{i\m)}_a&=&  V_{ij}^{(\r\m)}\bR^i_a
     + V_{ji}^{(\m\l}\pd_\l\bR^{i\r)}_a
     + \pd_\l(V_{ji}^{\l(\r}\bR^{i\m)}_a)\, ,             \nn\\
 0&=&\sum_{\pi(\r\m\l)}V_{ji}^{\r\m}\bR^{j\l}_a\, .        \lab{A.2}
\eea
As one can verify directly, these relations coincide with the
identities \eq{A.1}.

\section{The Noether form of \boldmath{$\tG^\m$}}

Here, we derive the formula \eq{4.6} for the canonical current
$\tG^\m$. Using the gauge transformations \eq{4.1}, the expression
\eq{3.5a} for $\tG^\m$ takes the form
\bea
\tG^\m&=& -2\left[ V_{ij}^\m\vphi^j +V_{ij}^{\m\n}\pd_\n\vphi^j
   +\pd_\n(\vphi^j V_{ji}^{\m\n})\right]
          (\bR^i_a\ve^a+\bR^{i\l}_a\pd_\l\ve^a)
   +2\pd_\n(\vphi^j V_{ji}^{\m\n}\bR^i)+\cO_2              \nn\\
&=&-2\left[ V_{ij}^\m\vphi^j +V_{ij}^{\m\n}\pd_\n\vphi^j
   +\pd_\n(\vphi^j V_{ji}^{\m\n})\right]\bR^i_a\ve^a       \nn\\
&&+2\pd_\l\left\{\left[ V_{ij}^\m\vphi^j
                       +V_{ij}^{\m\n}\pd_\n\vphi^j
  +\pd_\n(\vphi^j V_{ji}^{\m\n})\right]
                        \bR^{i\l}_a\right\}\ve^a           \nn\\
&&-2\pd_\l\left\{ \left[ V_{ij}^\m\vphi^j
     +V_{ij}^{\m\n}\pd_\n\vphi^j
     +\pd_\n(\vphi^j V_{ji}^{\m\n})\right]
 \bR^{i\l}_a\ve^a-\vphi^j V_{ji}^{\m\l}\bR^i\right\}+\cO_2\,,\nn
\eea
This result can be rewritten as
$$
\tG^\m=N^\m_a\ve^a+\pd_\r N^{\m\r}+\cO_2\, ,
$$
where
\bea
N^\m_a=&&2\left[ (V^\m_{ji}-\pd_\r V_{ji}^{\m\r})\bR^i_a
  -\pd_\l\left((V_{ji}^\m-\pd_\r V_{ji}^{\m\r})\bR^{i\l}_a\right)
           \right]\vphi^j                                   \nn\\
&&+2\left[-2V_{ij}^{(\m\n)}\bR^i_a
           -(V_{ji}^\m-\pd_\r V_{ji}^{\m\r})\bR^{i\n}_a
  +2\pd_\l(V_{ij}^{(\m\n)}\bR^{i\l}_a)\right]\pd_\n\vphi^j  \nn\\
&&+2\left[ 2V_{ij}^{(\m\n)}\bR^{i\l}_a \right]
               \pd_\n\pd_\l\vphi^j +\cO_2\, ,               \nn\\
N^{\m\l}=&& 2\left[(V_{ji}^\m-\pd_\r V_{ji}^{\m\r})\vphi^j
    -2V_{ij}^{(\m\n)}\pd_\n\vphi^j \right]\bR^{i\l}_a\ve^a  \nn\\
&&+2\left[V_{ji}^{\m\l}\vphi^j\right]
          (\bR^i_a\ve^a+\bR^{i\r}_a\pd_\r\ve^a)+\cO_2\, .   \nn
\eea
The form \eq{4.5} of the field equations implies
$N^\m_a\ve^a=F_i\bR^{i\m}_a\ve^a+\cO_2$, as we expected. Similarly,
using the identities derived in Appendix A, one finds that the above
expression for $N^{\m\l}$ can be brought to the form \eq{4.6b}.

\end{document}